\documentclass[letterpaper,twocolumn,pra,showpacs]{revtex4}% Physical Review B

\usepackage{graphicx}% Include figure files
\usepackage{dcolumn}% Align table columns on decimal point
\usepackage{bm}% bold math
\usepackage{epsfig}
\usepackage{amsmath,amsthm,amssymb}
\begin{document}

%\preprint{APS/123-QED}

\title{Cyclic Networks of Quantum Gates}

\author{Peter Cabauy$\,^{1,2,}$}
 \email{cabauy@umich.edu}
\author{Paul Benioff$\,^{1,}$}
 \email{pbenioff@anl.gov}
\affiliation{$^{1}$Physics Division, Argonne National Laboratory,
Argonne, IL  60439}
 \affiliation{$^{2}$Applied Physics Department,
University of Michigan, Ann Arbor, MI 48109}

\date{\today}% It is always \today, today,

\begin{abstract}
In this article initial steps in an analysis of cyclic networks of
quantum logic gates will be given.  Cyclic networks are those in
which the qubit lines are loops. In our investigations of cyclic
networks of quantum gates we have studied one and two qubit
systems plus two qubit systems connected to another qubit on an
acyclic line. The analysis includes: classifying networks into
groups, the dynamics of the qubits in the cyclic network, and the
perturbation effects of an acyclic qubit acting on a cyclic
network of quantum gates. This will be followed by a discussion on
quantum algorithms and quantum information processing with cyclic
networks of quantum gates, a novel implementation of a cyclic
network quantum memory and quantum sensors via cyclic networks
will also be discussed.
\end{abstract}

\maketitle
\section{Introduction}

To date, quantum information research has largely been concerned
with processing information in an acyclic manner.  This is due to
the fact that in 1993 Andrew Yao~\cite{turing:acyclic} showed that
any function computable in polynomial time by a quantum Turing
machine can also be computed by a polynomial-sized, acyclic
quantum gate array. This result showed that acyclic quantum gate
arrays are sufficient for modelling computations and consequently
in the years after Yao's result, quantum algorithms have been
described based on acyclic arrays.

Interestingly, some algorithms have a cyclic iterative component
built into their evolution yet they are still expressed by acyclic
arrays.  For instance, Grover's search algorithm, expressed as an
acyclic array, repeats a set of quantum gate operations
O($\sqrt{n}$) times to reach a solution \cite{grover}. These
repeating set of gates can just as easily be expressed as a single
set of quantum gates looped back onto it-self where the qubits are
measured after O($\sqrt{n}$) cycles around the network.  In other
words, the acyclic array for Grover's algorithm can be depicted
more compactly with a cyclic network.  Another iterative process
that can easily be expressed as a cyclic network is quantum phase
estimation \cite{Kit,revisited}.  In this algorithm, typically
represented by an acyclic array, a unitary operator is iterated
conditionally for O($2^t$) times (where $t$ is chosen to be the
desired bit length estimate of the phase).  This iterated operator
can also be expressed as a cyclic network in conjunction with an
acyclic line acting conditionally on the cyclic network (to be
shown in Section 4, Figure 9).

In realizing that these quantum algorithms have an iterative
property; yet are typically depicted by acyclic gate arrays, it
becomes interesting to study the structure and evolution of cyclic
networks of quantum gates.  One reason for this interest is that
cyclic networks are common place in ''classical" computing.  For
instance, the ''do loop" subroutine is often used in programs and
would be cumbersome to implement using acyclic methods.  Also,
computer hardware is comprised of wiring circuitry which involve
many loops.

Other reasons for studying cyclic networks of quantum gates stem
from the fact that some physical systems may require networks that
are compact with just a few inputs and outputs, and a great deal
of internal looping. For example, one can imagine a quantum robot
\cite{benioff-robot1,benioff-robot2} moving about an environment
(i.e. a lattice) where an on-board quantum computer controls the
robot's operations.  In this case, it is difficult to describe the
on board quantum computer with acyclic arrays given the limited
volume that a quantum robot encompasses.

Another reason for studying cyclic networks is that, unlike the
case for acyclic networks,  they are not limited to computations
that halt.  Halting computations can be carried out in cyclic
networks by periodic measurements of a flag qubit to determine if
the computation has halted.  Also it may not even be decidable
which acyclic array is equivalent to a cyclic network of arbitrary
complexity, or even if an equivalent acyclic array exists.  This
is based on the observation that the existence problem seems
equivalent to the unsolvable halting problem for Turing machines.

At the present time some research into the ''cyclic" processing of
quantum information within the context of feedback systems has
been looked at by Lloyd \cite{coherent}.  In his investigation, it
was shown that quantum systems benefit from information in a
feedback procedure where no measurement is made.  Rather, the
information is processed by a ''quantum governor" that may be
modelled by a quantum gate network; which in turn sends quantum
feedback information back to the quantum system, thus completing a
cyclic quantum gate procedure. Other investigations in quantum
feedback within the context of quantum control have also appeared
in the literature~\cite{viola, control1, control2}.

It is possible that an outright investigation into cyclic networks
of quantum gates may inspire new methods for designing quantum
algorithms and overall quantum information processing.  In our
investigations of cyclic networks, we have studied simple one and
two qubit systems.  Although the one qubit cyclic networks are
quite trivial, they will be used to motivate the analysis of the
less trivial two qubit cyclic networks.  In this article, cyclic
networks will be classified into groups (Section II), the dynamics
of the qubits moving around the cyclic networks will be examined
(Section III), and the perturbation effects of an acyclic qubit
acting on a cyclic network will be looked at (Section IV). A
discussion will follow on how cyclic networks might be used in
quantum algorithms and quantum information processing, a
description of a novel implementation of a cyclic network quantum
memory will be given, and quantum sensors via cyclic networks will
also be discussed.

\section{Structure of Simple Cyclic Networks}

\begin{figure}[t]
    \centering{\epsfig{file=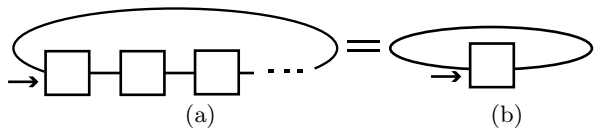}}
    \put(-155,-12){\makebox(0,0)[bl]{\small{(a)}\hspace{36.5mm}\small{(b)}}}
    \caption{(a)-A sequence of single qubit gates where each gate belongs to the U(2) group but the parameters of each gate
    may vary from gate to gate.  (b)-One gate with appropriate parameters in U(2) can simulate the cyclic network on the left
    hand side.}
    \label{SO}
\end{figure}

In order to explore simple cyclic networks it is useful to
classify them according to the structure of the unitary matrix
representing the cyclic network's combined gate-operation for one
iteration of the qubit(s). It is helpful to use the following two
conventions when classifying these networks into groups: The first
convention sets the number of qubits per line to only one. The
second convention requires the qubits to move in the direction of
the arrows (see Figures~\ref{SO}, \ref{twoqubit-gates}) and to
move simultaneously through both lines of any two qubit gate.

As will become evident later, one reason for classifying these
cyclic networks (restricted to the above two conventions) is to be
able to compactly express complicated gate arrangements within a
cyclic network with fewer gates. This will allow for an easy way
to understand the evolution of the qubit's wavefunction for
complicated gate arrangements by using a less complicated
arrangement that represents its group structure.

\subsection{One Qubit Structure}

As a simple introductory example to the classification process,
consider the one qubit cyclic network in Figure~\ref{SO}(a). A
cyclic network of this type may be simulated by a single qubit
gate acting on the $|0\rangle$, $|1\rangle$ basis with the general
U(2) group (Figure~\ref{SO}(b)).  In matrix form the well known
U(2) group~\cite{morton} may be represented by
\begin{equation}\label{U2}
  e^{i\delta}\begin{pmatrix} e^{i\alpha}cos\phi & e^{i\beta}sin\phi \\
  -e^{-i\beta}sin\phi & e^{-i\alpha}cos\phi \end{pmatrix}
\end{equation}

The cyclic network may be classified further if the single qubit
gate in Figure~\ref{SO}(b) is reduced to the general SU(2) group
by setting $\delta=0$.  In this case, the gates represented in
Figure~\ref{SO}(a) are restricted to SU(2) and its subgroups.
Similarly, if the single qubit gate is reduced to the SO(2) group
where $\alpha=\beta=\delta=0$, then the gates in
Figure~\ref{SO}(a) are restricted to the SO(2) group.

\subsection{Two Qubit Structure}
Gates with two interacting qubits have a more interesting
structure and can be similarly classified into U(2), SU(2) and
SO(2) groups. (For an overview of quantum gates see reference
\cite{elementary:gates, quantum:gates, quantum_computing_book,
ekert}.) The action of a general Control-U(2) gates may be defined
as
\begin{equation}\label{Gdn}
  G_{dn}\scriptstyle{(\alpha,\phi,\beta,\delta)}=
  \begin{pmatrix}
    1 & 0 & 0 & 0 \\
    0 & 1 & 0 & 0 \\
    0 & 0 & e^{i(\alpha+\delta)}cos\phi & e^{i(\beta+\delta)}sin\phi \\
    0 & 0 & -e^{-i(\beta-\delta)}sin\phi  & e^{i(-\alpha+\delta)}cos\phi
  \end{pmatrix}
 \end{equation}
where the U(2) matrix acts on the $|10\rangle,|11\rangle$ basis,
and
\begin{equation}\label{Gup}
 G_{up}\scriptstyle{(\alpha,\phi,\beta,\delta)}=
  \begin{pmatrix}
    1 & 0 & 0 & 0 \\
    0 & e^{i(\alpha+\delta)}cos\phi & 0 & e^{i(\beta+\delta)}sin\phi \\
    0 & 0 & 1 & 0 \\
    0 & -e^{-i(\beta-\delta)}sin\phi & 0  & e^{i(-\alpha+\delta)}cos\phi
  \end{pmatrix}
\end{equation}
where the U(2) matrix acts on $|01\rangle,|11\rangle$ basis.
(Note: schematically \vspace{.02in} $G_{dn}$ is represented by
$\;\;$
\begin{picture}(20,10)
    \put(5,-4){\framebox(6,6)}\put(8,2){\line(0,0){7}}\put(20,8.9){\line(-1,0){25}}
    \put(4,-2){\line(-1,0){9}}\put(21,-2){\line(-1,0){9}}
\end{picture}
\hspace{.02in} and $G_{up}\;$ is represented by $\;\;$
\begin{picture}(20,10)
    \put(5,3){\framebox(6,6)}\put(8,-4){\line(0,0){7}}
    \put(4,6){\line(-1,0){9}}\put(21,6){\line(-1,0){9}}
    \put(20,-4){\line(-1,0){25}}
\end{picture}
\hspace{.02in}) Both control gates act on a column vector
representing the binary basis in lexicographical order.
\begin{figure}[t]
    \centering{\epsfig{file=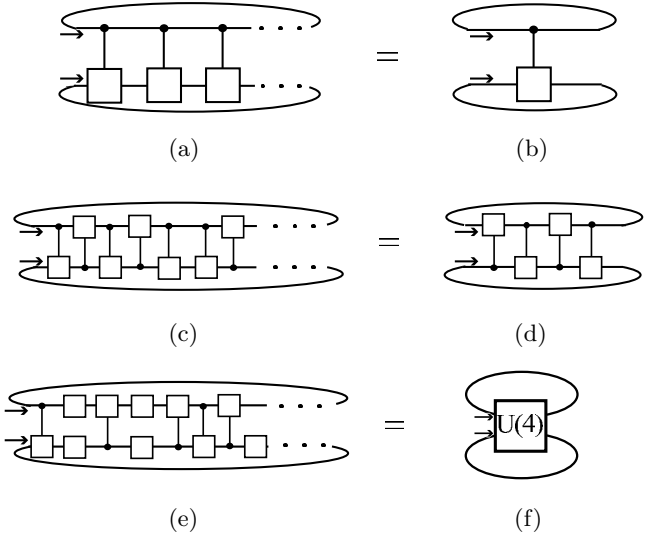,height=2.5in}}
    \put(-180,120){\makebox(0,0)[bl]{\small{(a)}\hspace{42mm}\small{(b)}}}
    \put(-180,50){\makebox(0,0)[bl]{\small{(c)}\hspace{42mm}\small{(d)}}}
    \put(-180,-20){\makebox(0,0)[bl]{\small{(e)}\hspace{42mm}\small{(f)}}}
    \caption{(a,c,e)-A sequence of gates making U(2) transformations on a two qubit cyclic network. Note, each gate may be
    a different U(2) transformation with different parameters in U(2).  (b,d,f)-A sequence of gates that can simulate the
    cyclic network on the left hand side with fewer gates.}
    \label{twoqubit-gates}
\end{figure}

In considering a product of gates $G_{dn}$ in
Figure~\ref{twoqubit-gates}(a) the product of U(2) matrices only
acts on the lower 2x2 matrix of $G_{dn}$.  Therefore, the
consecutive product of similarly oriented $G_{dn}$ gates is
equivalent to just one $G_{dn}$ gate as in
Figure~\ref{twoqubit-gates}(b). (Note, the network has the same
group properties as in the single qubit case and may again be
restricted to the SU(2) and SO(2) groups.) This analysis is
similarly true for the $G_{up}$ gates with the exception that the
U(2) portion of the $G_{up}$ matrix acts on the basis states
$|01\rangle,|11\rangle$ rather than $|10\rangle,|11\rangle$ for
the $G_{dn}$ case.

The ability to compress similarly oriented control gates as in
Figure~\ref{twoqubit-gates}(a) to just one control gate as in
Figure~\ref{twoqubit-gates}(b) raises the question of how many
control gates are needed to simulate any arrangement of
alternating gates as in Figure~\ref{twoqubit-gates}(c).  The
answer to this question can be found by looking at the matrix
structure of the alternating product of matrices $G_{up}$ and
$G_{dn}$ represented by
\begin{equation}\label{G}
  G=
  \begin{pmatrix}
    1 & 0 & 0 & 0 \\
    0 & m_{1\,1} & m_{1\,2} & m_{1\,3} \\
    0 & m_{2\,1} & m_{2\,2} & m_{2\,3} \\
    0 & m_{3\,1} & m_{3\,2} & m_{3\,3}
  \end{pmatrix}
\end{equation}
The product of assorted $G_{up}$ and $G_{dn}$ control gate
operations "actively" operate on three basis states
$|01\rangle,|10\rangle,|11\rangle$ with the lower 3x3 matrix which
may be called M. In contrast, the basis state $|00\rangle$ is only
multiplied by the identity operation.

Since the active operations are limited to the lower 3x3 matrix M,
it becomes evident that the most general 3x3 unitary matrix U(3)
will encompass all sequences of $G_{up}$ and $G_{dn}$ gate
operations. This matrix will be fairly complicated with nine
parameters making it difficult to find an arrangement of control
gates that will give M as a general U(3) matrix.

Fortunately, a prescription for a general U(3) matrix expressed as
a product of U(2) transformations exists~\cite{group1, group2}.
This approach considerably simplifies the process of finding the
control gate arrangement representing a general U(3) operation
because control gates are also U(2) transformations.

Following the prescription given in reference~\cite{group2} the
general U(3) matrix may be written as
\begin{equation}\label{U(3):1}
D(1,\gamma_{1},\gamma_{2},\gamma_{3})U_{3,4}(\phi_{1},\beta_{1})
  U_{2,3}(\phi_{2},\beta_{2})U_{2,4}(\phi_{3},\beta_{3})
\end{equation}
where $D(1,\gamma_{1},\gamma_{2},\gamma_{3})$ represents a
diagonal matrix with corresponding matrix entries
$1,e^{i\gamma_{1}},e^{i\gamma_{2}},e^{i\gamma_{3}}$. The
$U_{p,r}(\phi,\beta)$ matrix elements are obtained from a four
dimensional identity matrix with elements
$U_{p,p},U_{r,p},U_{p,r}$ and $U_{r,r}$ replaced with the
corresponding unitary matrix elements in Equation~\ref{U2} with
$\alpha,\delta=0$. The mapping of the elements is as follows:
\begin{equation}\label{Upq-matrices1}
U_{3,4}(\phi,\beta)=
  \begin{pmatrix}
    1 & 0 & 0 & 0 \\
    0 & 1 & 0 & 0 \\
    0 & 0 & cos\phi & e^{i\beta}sin\phi \\
    0 & 0 & -e^{-i\beta}sin\phi & cos\phi
  \end{pmatrix}
\end{equation}
\begin{equation}\label{Upq-matrices2}
U_{2,3}(\phi,\beta)=
  \begin{pmatrix}
    1 & 0 & 0 & 0 \\
    0 & cos\phi  & e^{i\beta}sin\phi  & 0 \\
    0 & -e^{-i\beta}sin\phi  & cos\phi & 0 \\
    0 & 0 & 0 & 1
  \end{pmatrix}
\end{equation}
\begin{equation}\label{Upq-matrices3}
U_{2,4}(\phi,\beta)=
  \begin{pmatrix}
    1 & 0 & 0 & 0 \\
    0 & cos\phi  & 0 & e^{i\beta}sin\phi\\
    0 & 0 & 1& 0 \\
    0 & -e^{-i\beta}sin\phi & 0 & cos\phi
  \end{pmatrix}
\end{equation}
(Note, that in this notation the general U(3) matrix appears as
the lower 3x3 matrix in order to conform with the typical matrix
structure for a product of $G_{up}$ and $G_{dn}$ gates as in
Equation~\ref{G}.)

A quick inspection of $U_{3,4}$ and $U_{2,4}$ matrices shows us
that these are nothing more than control gates $G_{up}$ and
$G_{dn}$ respectively.  However, $U_{23}$ does not have this form
but may be converted into gate form by realizing that
\begin{equation}\label{transf}
U_{2,3}(\phi_{2},\beta_{2})=U_{24}(-\frac{\pi}{2},0)U_{3,4}(\phi_{2},-\beta_{2})U_{24}(\frac{\pi}{2},0)
\end{equation}

Therefore, by including Equation~\ref{transf} into
Equation~\ref{U(3):1}, distributing the diagonal elements of
$D(1,\gamma_{1},\gamma_{2},\gamma_{3})$ and reducing similarly
oriented control gates to just one control gate, the U(3) matrix
is reduced to the following form of $G_{up}$ and $G_{dn}$ gates:
\begin{multline}\label{U(3):2}
U_{3,4}(\phi_{1},\beta_{1},\gamma_{2},\gamma_{3})U_{2,4}(-\frac{\pi}{2},0,\gamma_{1},0)\\
    \cdot U_{3,4}(\phi_{2},-\beta_{2},0,0)U_{2,4}(\phi_{3}+\frac{\pi}{2},\beta_{3},-\beta_{3},\beta_{3})
\end{multline}
Note, the $U_{p,r}$ matrices have been extended to have two more
variables where
\begin{equation}\label{4-variables}
  U_{2,4}(\phi,\beta,\gamma',\gamma'')=D(1,\gamma',1,\gamma'')U_{2,4}(\phi,\beta,0,0)
\end{equation}
and
\begin{equation}\label{4-variables2}
U_{3,4}(\phi,\beta,\gamma',\gamma'')=D(1,1,\gamma',\gamma'')U_{3,4}(\phi,\beta,0,0)
\end{equation}
where $U_{p,r}(\phi,\beta)=U_{p,r}(\phi,\beta,0,0)$.

Using this arrangement of matrices to give a general U(3)
operation with four alternating control gates, answers the
question of how many control gates are needed to simulate any
number or sequence of control gates.  This can be represented
pictorially as in Figure~\ref{twoqubit-gates}(d).

The question arises regarding how many control gates are needed to
simulate any alternating arrangement of Control-SU(2) and
Control-SO(2) gates with at least one Control-SU(2) gate present.
It is seen that four control gates is sufficient. This is due to
the fact that an arrangement of this type will only restrict the
matrix M to an SU(3) matrix which is the same as requiring
\begin{equation}\label{SU3-restriction}
\delta_{1}+\delta_{2}+\delta_{3}=0
\end{equation}
This on its own is not sufficient to reduce the number of control
gates.

On the other hand, if the cyclic network in
Figure~\ref{twoqubit-gates}(c) is limited to only Control-SO(2)
gates, then the number of gates required to simulate any
arrangement is three.  In fact, there is a very simple visual
interpretation for this result.  Consider the 3 basis states that
M operates on $|01\rangle,|10\rangle,|11\rangle$ to be the 3
dimensional cartesian axes (x,y,z).  An application $G_{dn}(\phi)$
on the two qubits leaves the coefficient of the basis state
$|00\rangle$ unchanged. However, it rotates a vector representing
the qubit-wavefunction about the $|01\rangle$ state (x-axis).
Similarly, an application of $G_{up}(\phi)$ is nothing more than a
rotation about the $|10\rangle$ state (y-axis). Given this, it is
well known that 3 rotations about alternating axes (i.e x and y
axis) give any rotation in 3 dimensional space or the SO(3)
group~\cite{goldstein}. Therefore, an application of three
alternating gates such as
\begin{equation}\label{SO(3)}
  G_{up}(\phi_{1})G_{dn}(\phi_{2})G_{up}(\phi_{3})=R_{y}(\phi_{1})R_{x}(\phi_{2})R_{y}(\phi_{3})
\end{equation}
\begin{equation}\label{SO(3):2}
  G_{dn}(\phi_{1})G_{up}(\phi_{2})G_{dn}(\phi_{3})=R_{x}(\phi_{1})R_{y}(\phi_{2})R_{x}(\phi_{3})
\end{equation}
gives a general operation in SO(3), ultimately showing that three
alternating gates is sufficient to simulate any number or
arrangement of Control-SO(2) gates.

Up to this point, the two qubit cyclic networks have been
classified into groups varying from U(3) through SO(2) control
gate arrangements.  This however does not encompass the most
general two qubit network where single qubit gates are used along
with control gates.  The most general two qubit network is the
U(4) group and it can encompasses any arrangement of single qubit
gates and two qubit control gates as depicted in
Figure~\ref{twoqubit-gates}(e).

A general U(4) matrix represented by a product of U(2)
transformations can again be found in reference~\cite{group2} and
is given by
\begin{multline}
    D(\gamma_{1},\gamma_{2},\gamma_{3},\gamma_{4})
        U_{3\:4}(\phi_{1},\theta_{1})\\
        \cdot U_{2\:3}(\phi_{2},\theta_{2})
        U_{2\:4}(\phi_{3},\theta_{3})
        U_{1\:2}(\phi_{4},\theta_{4})\\
        \cdot U_{1\:3}(\phi_{5},\theta_{5})
        U_{1\:4}(\phi_{6},\theta_{6})
\end{multline}
Each of these matrices can be converted into gates as follows
(Note, the labelling of the boxes denote the type of U(2)
operation to be implemented  by either a single qubit gate or
"conditionally" by a control gate.)
\begin{equation}\label{U4-bunchofgates:1}
    D(\gamma_{1},\gamma_{2},\gamma_{3},\gamma_{4})=\epsfig{bbllx=-10,bblly=20,bburx=100,bbury=50,file=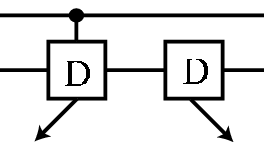}
    \put(-60,-30){\makebox(0,0)[b]{$\scriptstyle{D=D(\delta_{3},\delta_{4})\hspace{5mm}D=D(\gamma_{1},\gamma_{2})}$}}
\end{equation}
where $\gamma_{1}+\delta_{3}=\gamma_{3}$ and
$\gamma_{2}+\delta_{4}=\gamma_{4}$.
\begin{equation}\label{U4-bunchofgates:2}
    U_{3\:4}(\phi_{1},\theta_{1})=\epsfig{bbllx=-10,bblly=20,bburx=90,bbury=47,file=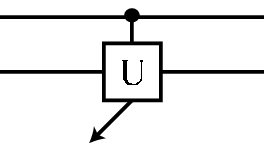}
    \put(-60,-30){\makebox(0,0)[b]{$\scriptstyle{U=U(\phi_{1},\theta_{1})}$}}
\end{equation}
\begin{equation}\label{U4-bunchofgates:3}
    U_{2\:3}(\phi_{2},\theta_{2})=\epsfig{bbllx=-10,bblly=20,bburx=90,bbury=65,file=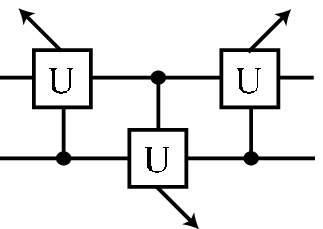}
    \put(-100,46){\makebox(0,0)[bl]{$\scriptstyle{U=U(\frac{\pi}{2},0)\hspace{15mm}U=U(-\frac{\pi}{2},0)}$}}
    \put(-25,-30){\makebox(0,0)[b]{$\scriptstyle{U=U(\phi_{2},-\theta_{2})}$}}
\end{equation}
\begin{equation}\label{U4-bunchofgates:4}
    U_{2\:4}(\phi_{3},\theta_{3})=\epsfig{bbllx=-10,bblly=2,bburx=90,bbury=46,file=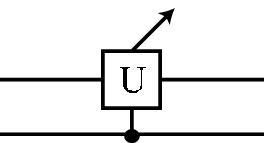}
    \put(-26,39){\makebox(0,0)[b]{$\scriptstyle{U=U(\phi_{3},\theta_{3})}$}}
\end{equation}
\begin{equation}\label{U4-bunchofgates:5}
    U_{1\:2}(\phi_{4},\theta_{4})=\epsfig{bbllx=-10,bblly=19,bburx=90,bbury=56,file=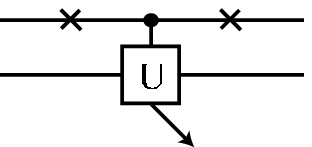}
    \put(-25,-28){\makebox(0,0)[b]{$\scriptstyle{U=U(\phi_{4},\theta_{4})}$}}
\end{equation}
The
\begin{picture}(10,10)
    \put(2,-1){\epsfig{file=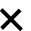}}
\end{picture}
represents the Not operation,
\begin{equation}\label{U4-bunchofgates:6}
    U_{1\:3}(\phi_{5},\theta_{5})=\epsfig{bbllx=-10,bblly=3,bburx=90,bbury=40,file=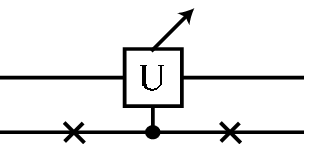}
    \put(-25,42){\makebox(0,0){$\scriptstyle{U=U(\phi_{5},\theta_{5})}$}}
\end{equation}
\begin{equation}\label{U4-bunchofgates:7}\sqrt{}
    U_{1\:4}(\phi_{6},\theta_{6})=\epsfig{bbllx=-5,bblly=20,bburx=140,bbury=55,file=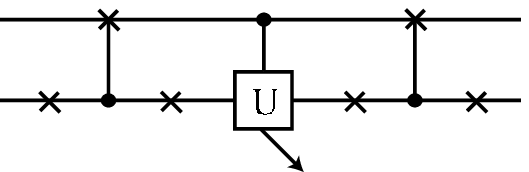,height=.45in}\\
    \put(-62,-28){\makebox(0,0)[bl]{$\scriptstyle{U=U(\phi_{6},\theta_{6})}$}}
\end{equation}
and the two qubit gate with the
\begin{picture}(10,10)
    \put(2,-1){\epsfig{file=control-notcompr.eps}}
\end{picture}
is the Control-Not gate.

Therefore, by connecting these gates end to end (in reverse order
due to the direction of the cycling qubits) a general U(4) cyclic
network is achieved and may be represented by the two qubit U(4)
box in Figure~\ref{twoqubit-gates}(f). Furthermore, by using the
grouping of single qubit gates and control gates discussed thus
far, it is possible to reduce the number of gates in the U(4)
network to just eight control gates and five single qubit gates.
Whether or not this is the smallest number of gates needed to
represent a general U(4) group remains to be investigated.

Interestingly, by using this network of gates represented by the
U(4) box it is possible to get the most general 3 qubit network
(U(8) group). The arrangement for this was given and discussed by
Barenco et.al~\cite{elementary:gates}.  The most general q qubit
network ($U(2^{q})$ group) can be achieved in a similar manner.
However, the number of U(4) networks will grow exponentially with
number of qubit lines q.

In all, cyclic quantum gate networks may be classified according
to the structure of the unitary matrix representing the cyclic
network's combined gate-operations for one iteration of the
qubit(s).  This is helpful when analyzing the evolution of simple
cyclic networks because understanding the evolution of one group
encompasses many different types of cyclic gate arrangements
within that group.  In the next section, the classification of
these simple groups will be quite useful because it simplifies the
eigenvalue and eigenstate analysis of cyclic networks to just
those representing the most general group arrangements.

\section{Eigenvalues and Eigenstates}
One of the goals in studying cyclic quantum networks is to
understand the evolution of the wavefunction for qubits cycling
through the network. For example, the evolution of a cyclic
network, such as in Figures \ref{SO}-\ref{twoqubit-gates}, take
the form
\begin{equation}\label{iteration}
  \Psi(n)=(U)^{n}\Psi(0)
\end{equation}
where $n$ is the number of cycles completed and $U$ represents the
gate operations resulting from one cycle around the network.

In practice, it is convenient to express Equation \ref{iteration}
in the $U$ operator eigenbasis.  The expansion of $\Psi$ in the
$U$ eigenbasis will take the form
\begin{equation}\label{psi:eigenbasis}
  \Psi(0)=\sum_{\mu=1}^{2^q} c_{\mu} \Psi_{\mu}
\end{equation}
where q is the number of qubit loops and $\Psi_{\mu}$ is the
eigenstate corresponding to eigenvalue $e^{i\nu_{\mu}}$.

Use of Equation~\ref{psi:eigenbasis} shows that
\begin{equation}\label{psi:evolution-eigenbasis}
\Psi(n)=\sum_{\mu=1}^{2^q} c_{\mu} e^{in \nu_{\mu}}\Psi_{\mu}
\end{equation}

\subsection{One Qubit Eigenvalues and Eigenstates}

>From the classification of cyclic networks in the previous chapter
it is understood that a general U(2) cyclic network with the
matrix operator given by Equation~\ref{U2} gives the U$(2)$
network's evolution, in the eigenbasis as
\begin{equation}
  U^{n}\Psi=c_{1} e^{in\nu_{1}}\Psi_{1} + c_{2} e^{-in\nu_{1}}\Psi_{2}
\end{equation}
where $\nu_{2}=-\nu_{1}$.  This also represents the evolution of
SU(2) and SO(2) networks with appropriate restrictions of
$\alpha,\beta,\delta$.

\subsection{Two Qubit Eigenvalues and Eigenstates}

As discussed in section II the most general two qubit network
belongs to the U(4) group. It operates on all four basis states
$|00\rangle,|01\rangle,|10\rangle,|11\rangle$, with a set of
control gates and single qubit gates as shown in Equations
\ref{U4-bunchofgates:1}-\ref{U4-bunchofgates:7}.  Since the main
interest here is in U(3), SU(3) and SO(3) networks, the discussion
of the evolution of these networks will be limited to these
groups.

\subsubsection{Eigenvalues for U(3) and Subgroups}

The evolution of a cyclic network belonging to U(3) or one of its
subgroups is given by Equation \ref{psi:evolution-eigenbasis}.
Replacing $U$ by $G$ from Equation~\ref{G} gives the following for
$\Psi(n)=G^{n}\Psi$:
\begin{equation}\label{arbitraryPsi:eigenbasis3}
 G^{n} \Psi=c_{1} e^{in\nu_{1}} \Psi_{1} + c_{2} e^{in\nu_{2}} \Psi_{2}
  + c_{3} e^{in\nu_{3}} \Psi_{3}
  + c_{4} e^{in\nu_{4}} \Psi_{4}
\end{equation}

The 4 eigenvalues for G are $\lambda = 1$ and 3 eigenvalues
corresponding to the following characteristic equation of degree
three
\begin{equation}\label{cubicequation}
\lambda^{3}+a_{1}\lambda^{2}+a_{2}\lambda+a_{3}=0
\end{equation}
The $a_{s}$ represent the principal minors \cite{schaums} of the
lower 3x3 matrix M in G, Equation~\ref{G}.

The solution to this general cubic equation with real or complex
coefficients may be found in references \cite{algebra, eric}, and
is given to be

\begin{equation}\label{cubic-solution}
\lambda_{k}=w^{\frac{1}{3}}e^{\frac{ik2\pi}{3}}-\frac{P}{3w^{\frac{1}{3}}e^{\frac{ik2\pi}{3}}}-\frac{a_{1}}{3}\qquad
For \;\;k=0,1,2
\end{equation}
where

\begin{equation}\label{w}
  w=\frac{Q}{2}+\sqrt{\frac{Q^2}{4}+\frac{P^3}{27}}
\end{equation}
and
\begin{equation}\label{QP}
Q=\frac{9a_{1}a_{2}-27a_{3}-2a_{1}^3}{27}\;\;\;\;\;;\;\;\;\;
P=\frac{3a_{2}-a_{1}^2}{3}
\end{equation}

The coefficients $a_{1},a_{2},a_{3}$ in the cubic equation are
given by

\begin{align}\label{coeff}
a_{1}&=-TrM\notag\\
a_{2}&=M_{11}+M_{22}+M_{33}\\
a_{3}&=-detM\notag
\end{align}
where
\begin{align}\label{cofactors:1}
M_{11}&=det\begin{pmatrix}
     m_{22} & m_{23}\\
     m_{32} & m_{33}
   \end{pmatrix} \notag \\
   M_{22}&=det\begin{pmatrix}
     m_{11} & m_{13}\\
     m_{31} & m_{33}
     \end{pmatrix}\\
   M_{33}&=det\begin{pmatrix}
     m_{11} & m_{12}\\
     m_{21} & m_{22}
   \end{pmatrix} \notag
\end{align}

If we restrict the 2 qubit cyclic network from U(3) to an SU(3) by
setting $\delta_{1}+\delta_{2}+\delta_{3}=0$ the coefficients
$a_{1},a_{2},a_{3}$ can be immediately simplified to
\begin{align}\label{ai}
a_{1}&=-TrM\notag\\
a_{2}&=(TrM)^{*}\\
a_{3}&=-1\notag
\end{align}

The result for $a_{3}$ follows directly from the determinant 1
group property for the SU(3) group.  However, the reduction
$a_{2}=(TrM)^{*}$ is not as apparent. For this result it is
helpful to realize that the cubic equation
(Equation~\ref{cubicequation}) may be rewritten as
\begin{equation}\label{cubicequation1}
  (\lambda-\lambda_{0})(\lambda-\lambda_{1})(\lambda-\lambda_{2})=0
\end{equation}
giving the following relations for the coefficients $a_{s}$:
\begin{align}
&\lambda_{0}+\lambda_{1}+\lambda_{2}=-a_{1}\notag\\
&\lambda_{0}\lambda_{1}+\lambda_{1}\lambda_{2}+\lambda_{0}\lambda_{2}=a_{2}\\
&\lambda_{0}\lambda_{1}\lambda_{2}=-a_{3}\notag
\end{align}
By noting that $\lambda_{0}\lambda_{1}\lambda_{2}=1$ for the SU(3)
matrix and that the eigenvalues to the control gate networks are
of the form $\lambda_{k}=e^{i\nu_{k}}$ as for all unitary
matrices.  It becomes evident that
\begin{equation}\label{lambdarelation}
\lambda_{0}\lambda_{1}+\lambda_{1}\lambda_{2}+\lambda_{0}\lambda_{2}=(\lambda_{0}^{*}+\lambda_{1}^{*}+\lambda_{2}^{*})
\end{equation}
showing that $a_{2}=(TrM)^{*}$.

\begin{figure}
    \centering{\epsfig{file=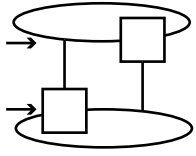}}
    \caption{An example of an SU(3) cyclic network with two alternating control gates of the form $G_{dn}(\alpha,
    \phi,\beta)$ and $G_{up}(\alpha,\phi,\beta)$.}
    \label{chap3Fig2}
\end{figure}

A particular example of a cyclic network belonging to the $SU(3)$
group is the network with an alternating pair of gates
$G_{_{up}}(\alpha,\phi,\beta)G_{_{dn}}(\alpha,\phi,\beta)$ as in
Figure \ref{chap3Fig2}. In this case, $TrM=A(\alpha,\phi)$ where
\begin{equation}\label{A}
  A(\alpha,\phi)=e^{-2i\alpha}cos^{2}(\phi) +
  2e^{i\alpha}cos(\phi).
\end{equation}
>From Equation~\ref{ai} one has $a_{1}=-A(\alpha,\phi)$ and
$a_{2}=A^{*}(\alpha,\phi)$. Therefore, the eigenvalues are
$\lambda=1$ and three other eigenvalues as the solutions to the
cubic eigenvalue equation
\begin{equation}\label{SU3:example}
\lambda^{3}-A(\alpha,\phi)\lambda^{2}+A^{*}(\alpha,\phi)\lambda-1=0
\end{equation}

The solutions to the cubic equation follow from Equation
\ref{cubic-solution} but can be simplified further by making use
of the property $A(\alpha\pm\frac{2\pi}{3},\phi)=
A(\alpha,\phi)e^{\frac{\pm i2\pi}{3}}$.  This can be verified
directly by substituting $\alpha \pm \frac{2\pi}{3}$ into
Equation~\ref{A}.  By making use of this property $w, Q, P$ from
Equations~\ref{w}-\ref{QP} have the following relations:
\begin{align}\label{w-P-Q}
  &w(\alpha\pm\frac{2\pi}{3},\phi)=w(\alpha,\phi)\notag\\
  &Q(\alpha\pm\frac{2\pi}{3},\phi)=Q(\alpha,\phi)\\
  &P(\alpha\pm\frac{2\pi}{3},\phi)=P(\alpha,\phi)e^{\frac{\mp
  i2\pi}{3}}\notag
\end{align}
and the solution to the cubic equation
(Equation~\ref{SU3:example}) takes on the simple form:
\begin{equation}\label{SU3:cubic-example:1}
  \lambda_{k}=\lambda_{0}(\alpha_{k},\phi)e^{\frac{ik2\pi}{3}}
\end{equation}
where
\begin{equation}\label{SU3:cubic-example:2}
  \lambda_{0}(\alpha_{k},\phi)=w^{\frac{1}{3}}(\alpha_{k},\phi) -
  \frac{p(\alpha_{k},\phi)}{3w^\frac{1}{3}(\alpha_{k},\phi)}+\frac{A(\alpha_{k},\phi)}{3}
\end{equation}
and
\begin{equation}\label{alpha-n}
  \alpha_{k}=\alpha-\frac{2k\pi}{3}
\end{equation}
As an example consider when $\alpha=\frac{\pi}{4}$, for this case
the three eigenvalue solutions using
Equation~\ref{SU3:cubic-example:1} are
$\lambda_{0}=\lambda_{0}(\frac{\pi}{4},\phi)$,
$\lambda_{1}=\lambda_{0}(\frac{\pi}{4}-\frac{2\pi}{3},\phi)e^{\frac{i2\pi}{3}}$
and
$\lambda_{2}=\lambda_{0}(\frac{\pi}{4}-\frac{4\pi}{3},\phi)e^{\frac{i4\pi}{3}}$.

\begin{figure}[t]
    \centering{\epsfig{file=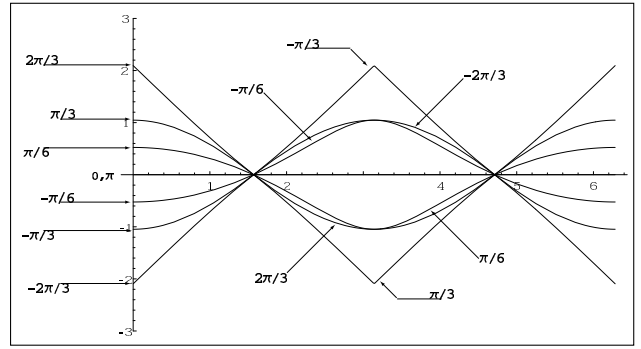,height=1.8in}}
    \caption{A plot of eigenphase $\nu_{0}$ corresponding to the eigenvalue $\lambda_{0}=e^{i\nu_{0}}$
    where the ordinate represents the
value of $\nu_{0}$, the abscissa represents the value of $\phi$
and each of the family of curves in the plot are given for
different $\alpha$.}
    \label{chap3Fig3}
\end{figure}

Since the three eigenvalues may be extracted from the eigenvalue
$\lambda_{0}$, it is sufficient to plot the eigenphase $\nu_{0}$
in order to analyze the properties of the three eigenvalues. A
plot of eigenphase $\nu_{0}$ corresponding to the eigenvalue
$\lambda_{0}=e^{i\nu_{0}}$ can be seen in Figure~\ref{chap3Fig3}
where the ordinate represents the value of $\nu_{0}$, the abscissa
represents the value of $\phi$ and each of the family of curves in
the plot are given for different $\alpha$.

There are some important properties that can be noted about these
eigenvalue solutions. The first property pertains to three special
solutions that are constant for all values of $\phi$. These are
the solutions for which the eigenphase $\nu=\alpha$ and can be
attained by directly substituting $\lambda=e^{i\alpha}$ into
Equation~\ref{SU3:example} and expanding out the $A(\alpha,\phi)$
as given by Equation~\ref{A}.  This gives the following equation
\begin{equation}\label{nu-alpha}
  (e^{3i\alpha}-1)(cos^{2}(\phi)-2cos(\phi)-1)=0
\end{equation}
with three special solutions
$\alpha=0,\frac{2\pi}{3},\frac{-2\pi}{3}$ corresponding to
$\lambda=1,e^{\frac{i2\pi}{3}},e^{-\frac{i2\pi}{3}}$. (Note, that
only one of these solutions appears in the plot and corresponds to
$\lambda_{0}$ with $\nu_{0}=0$.  The other two solutions are just
a shift as described by Equation~\ref{SU3:cubic-example:1}.)  A
second property is the symmetry
$\lambda_{0}(-\alpha)=\lambda_{0}(\alpha)^{*}$ which can be seen
in Figure~\ref{chap3Fig3} as the eigenphase $\nu_{0}$ is symmetric
about the $\nu_{0}=0$ axis. This can also be verified directly in
Equation~\ref{SU3:cubic-example:2} by noticing that $p$, $w$ have
this symmetry due to the fact that
$A(-\alpha,\phi)=A^{*}(\alpha,\phi)$.  Similarly, a third property
associated with the eigenvalues of this particular SU(3) network
is the translational symmetry
$\lambda(\alpha+\pi,\phi)=\lambda(\alpha,\phi+\pi)$ which can also
be directly associated to $A(\alpha,\phi)$ having the same
translational symmetry.  As a final comment on the $\nu_{0}$ plot,
the nodes appearing at the values $\phi=\frac{\pi}{2}$ and
$\phi=\frac{3\pi}{2}$ occur because $A(\alpha,\frac{\pi}{2})=0$
and $A(\alpha,\frac{3\pi}{2})=0$ for all values of $\alpha$.  A
simple examination of Equation~\ref{SU3:example} shows that for
these two values of $\phi$, $\lambda_{0}=1$ for all $\alpha$.

An interesting variation on this example is when $-\alpha$ appears
in one of the two alternating gates such as in the case
$G_{_{up}}(-\alpha,\phi,\beta)G_{_{dn}}(\alpha,\phi,\beta)$. The
trace M of this network now becomes
\begin{equation}
TrM=2cos(\alpha)cos(\phi) + cos^{2}(\phi)
\end{equation}
and the coefficients to the cubic equation are now real and take
on the relationship $a_{2}=-a_{1}$. Cases like this where the
trace is real simplify the solution to the form
\begin{equation}\label{evalues}
  \lambda_{0}=1,\quad
\lambda_{1,2}=\frac{TrM-1}{2}\pm\frac{i\sqrt{(3-TrM)(TrM+1)}}{2}
\end{equation}
where
\begin{equation}\label{evalues:2}
  cos\nu=\frac{TrM-1}{2}, \quad sin\nu=\frac{\sqrt{(3-TrM)(TrM+1)}}{2}
\end{equation}

Other cases where the network's trace is real belong to cyclic
networks with just a single Control-SU(2) or Control-SO(2) gate as
in Figure~\ref{twoqubit-gates}(b). This of course is trivial, and
for that matter the eigenvalues are the same as that in the single
qubit case. Of more interest is the general $SO(3)$ group.  In
this case, the three gates are $
G_{_{up}}(\phi_{1})G_{_{dn}}(\phi_{2})G_{_{up}}(\phi_{3})$ and the
trace of the network has the simple form
\begin{equation}
TrM=cos(\phi_{2})+cos(\phi_{1}+\phi_{3})(1+cos(\phi_{2}))
\end{equation}
In the next section a cyclic control gate network belonging to the
$SO(3)$ group will be used as an example to understand the
perturbative effects of an acyclic line acting on the cyclic
network.

\subsubsection{Eigenstates of U(3) and Subgroups} It is fairly
simple to write down the 4 unnormalized qubit eigenstates for
control gate cyclic networks.  The following unnormalized
eigenstates can be applied to any network with matrix-gate
structure G (Note: G matrix structure includes U(3) and subgroups)
\begin{gather}
    \Psi_{k}=
    \begin{pmatrix}
        0\\
        -m_{1\,3}(m_{2\:2}-\lambda_{k})+m_{1\,2}m_{2\,3}\\
        -m_{2\,3}(m_{1\,1}-\lambda_{k})+m_{2\,1}m_{1\,3}\\
        (m_{2\,2}-\lambda_{k})(m_{1\,1}-\lambda_{k})-m_{2\,1}m_{1\,2}
    \end{pmatrix}
    \begin{matrix}
        |00\rangle\\|01\rangle\\|10\rangle\\|11\rangle
    \end{matrix} \notag \\
    \quad \Psi_{3}=
    \begin{pmatrix}\label{eigenstatesG}
        1\\0\\0\\0
    \end{pmatrix}
    \begin{matrix}
        |00\rangle\\|01\rangle\\|10\rangle\\|11\rangle
    \end{matrix}
\end{gather}
$\Psi_{k}$ corresponds to eigenstates $k=0,1,2$ and $\lambda_{k}$
are the eigenvalues associated with the respective eigenstates.

\section{Perturbation of a Cyclic Network}
%do introduction after redoing chapter intro. after finishing the rest of the thesis
Some areas of investigation that may lead to potential
applications for cyclic quantum networks are quantum memories,
quantum sensors and the ability to make existing acyclic quantum
gate arrays more compact. As a first step in understanding the
feasibility of these potential applications, it is essential to be
able to interact with the cyclic network via qubit(s) on an
acyclic line. A very straightforward method of interacting with a
cyclic network is to connect the cyclic network to an acyclic
qubit via a control gate as seen in Figure~\ref{a}(a,b).

In the rest of this section an analysis of a general two qubit
cyclic network perturbed by an acyclic qubit will be examined.
Following this general description, a specific example will be
presented. (Note: The analysis of a perturbed single qubit cyclic
network can be readily achieved by following the outlined method
for the perturbed two qubit case.)

\subsection{Perturbation of a Two Qubit Cyclic Network}

\begin{figure}
    \centering{\epsfig{file=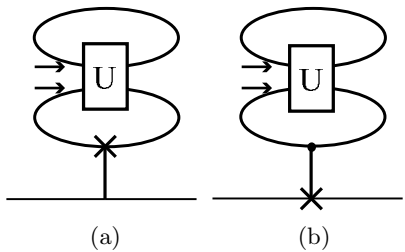}}
    \put(-119,-15){\makebox(0,0)[bl]{\small{(a)}\hspace{23.5mm}\small{(b)}}}
    \caption{Two qubit cyclic network perturbed by an acyclic qubit via a Control-Not gate.  Note, the box
    "U" denotes any set of gates acting in the 4 dimensional Hilbert Space.}
    \label{a}
\end{figure}

In examining the two qubit cyclic network in Figure~\ref{a}(a,b),
the acyclic line is connected to the bottom cyclic loop via a
Control-Not gate.  Box "U" in the cyclic network denotes any
arrangement of quantum gates, to include U(4) and its subgroups.
Prior to an interaction with the acyclic qubit, the cyclic network
evolves as previously described in section~III. However, after an
interaction the cyclic network evolves in a perturbed/entangled
state with the acyclic qubit. (Note: the acyclic qubit, and the
bottom, cyclic qubit must be coincident at the Control-Not gate at
the time of interaction). The acyclic qubit, which remains in an
entangled state with the cyclic network, proceeds past the
interaction gate as the cyclic qubits continue to move around the
network.

To examine this more closely, the following requirements are to be
considered. First, the state of the qubits in the cyclic network
is initially set to an arbitrary superposition
\begin{equation}\label{super}
\Psi=c_{1}|00\rangle+c_{2}|01\rangle+c_{3}|10\rangle+c_{4}|11\rangle
\end{equation}
Second, the cyclic network is allowed to evolve an arbitrary
number of $n$ iterations before it interacts with the acyclic
qubit. Third, an acyclic qubit in the arbitrary state
\begin{equation}\label{Phi}
\Phi=\alpha|0\rangle+\beta|1\rangle
\end{equation}
interacts with the cyclic network via a Control-Not gate, and then
continues its course on the acyclic line as the qubits in the
cyclic network continues to evolve for $n'$ more iterations.  In
considering a case like this the resulting state is
\begin{equation}\label{1qubitpert:first}
  \{P_{0} \otimes U^{(n'+n)} + P_{1} \otimes U^{n'}\sigma_{x,b}
U^{n}\}|\Phi \otimes \Psi\rangle
\end{equation}
for Figure~\ref{a}(a).  To write this equation the relation
$(1\otimes U)P_{1}\otimes\sigma_{x,b}=P_{1}\otimes U\sigma_{x,b}$
has been used.  The $P_{0}=|0\rangle\langle0|$ and
$P_{1}=|1\rangle\langle1|$ are projection operators on the state
of the perturbing qubit, $\sigma_{x}$ represents a Pauli Matrix
and the operator U is an arbitrary matrix which represents the
gate operation of the cyclic network upon one iteration. Note,
that the subscript b associated with the operators represent the
bottom qubit of the cyclic network.

If the Control-Not gate connected to the acyclic line is flipped
around so as to have the target bit on the acyclic line and the
control on the cyclic line as in Figure~\ref{a}(b), the resulting
state evolution then becomes
\begin{equation}\label{persecond}
\{1\otimes U^{n'}P_{0,b}U^{n}+\sigma_{x} \otimes
  U^{n'}P_{1,b}U^{n}\}|\Phi\otimes\Psi\rangle.
\end{equation}
To write this equation the relation $U\sigma_{x}\otimes
P_{1,b}=\sigma_{x}\otimes UP_{1,b}$ has been used. One aspect to
note about this analysis is that the Control-Not gate may easily
be replaced by a more general Control-U(2) gate by simply changing
the $\sigma_{x}$ for the more general U(2) operator.

\begin{table*}[t]
    \begin{center}
        \begin{tabular}{l|c|c|c} \hline
        \emph{$m_{j,j'}$}&
        \emph{$A(\phi)$}&
        \emph{$B(\phi)$}&
        \emph{$C(\phi)$}\\ \hline \hline
        $m_{1,1}$ & $\frac{(c+1)}{(c+3)}$ & $\frac{2}{(c+3)}$ & $\scriptstyle{0}$ \\ \hline

        $m_{1,2}$ & $\frac{-(c+1)}{(c+3)}$ &$\frac{4c-2c^{2}
             \cos\nu - 2 \cos\nu_{1}}{s^{2}(c+1)(c+3)}$
            & $\frac{-2\sin\nu_{1}}{(c+1)(c+3)}$ \\ \hline

        $m_{1,3}$ & $\frac{-s}{(c+3)}$ & $\frac{ 2c^{3} \cos\nu_{1}
            -6c^{2}+6c \cos\nu_{1} -2 \cos2\nu_{1}}{s^{3} (c+1)(c+3)}$
            & $\frac{-2c^{3} \sin\nu_{1} + 6c \sin\nu_{1} - 2 \sin2\nu_{1}}{s^{3}
            (c+1)(c+3)}$ \\ \hline

        $m_{2,1}$ & $\frac{-(c+1)}{(c+3)}$ & $\frac{4c-2c^{2}
            \cos\nu_{1} - 2 \cos\nu_{1}}{s^{2}(c+1)(c+3)}$ & $\frac{2
            \sin\nu_{1}}{(c+1)(c+3)}$ \\ \hline

        $m_{2,2}$ & $\frac{(c+1)}{(c+3)}$ & $\frac{2}{(c+3)}$ & $\scriptstyle{0}$ \\ \hline

        $m_{2,3}$ & $\frac{s}{(c+3)}$ & $\frac{2(-c^{3}+3c^{2}
            \cos\nu_{1} - c (\cos2\nu_{1} +2) + \cos\nu_{1})}{s^{3}(c+1)(c+3)}$
            & $\frac{2(c^{2} \sin\nu_{1} - c \sin2\nu_{1} +
            \sin\nu_{1})}{s^{3}(c+1)(c+3)}$ \\ \hline

        $m_{3,1}$ & $\frac{-s}{(c+3)}$ & $\frac{2c^{3} \cos\nu_{1} -
            6c^{2} + 6c \cos\nu_{1} - 2 \cos2\nu_{1}}{s^{3}(c+1)(c+3)}$
            & $\frac{2c^{3} \sin\nu_{1} - 6c \sin\nu_{1} + 2
            \sin2\nu_{1}}{s^{3}(c+1)(c+3)}$ \\ \hline

        $m_{3,2}$ & $\frac{s}{(c+3)}$ & $\frac{2(-c^{3} + 3c^{2}
            \cos\nu_{1} - c(\cos2\nu_{1} + 2) + \cos\nu_{1})}{s^{3}(c+1)(c+3)}$
            & $\frac{2(-c^{2} \sin\nu_{1} + c \sin2\nu_{1} -
            \sin\nu_{1})}{s^{3}(c+1)(c+3)}$ \\ \hline

        $m_{3,3}$ & $\frac{(1-c)}{(c+3)}$ & $\frac{2(c+1)}{(c+3)}$ &
            $\scriptstyle{0}$  \\ \hline

        \end{tabular}
        \caption{Coefficients $A(\phi),B(\phi),C(\phi)$ for matrix elements $m_{j,j'}$ in $G^{n}$. Note, $c=\cos\phi$ and $s=\sin\phi$.}
        \label{mjj}
    \end{center}
\end{table*}

The matrix elements of $U^{n}$ in the binary basis are given by
\begin{equation} \label{matrix:n}
\langle j'|U^{n}|j \rangle=\sum_{\mu}
  |j'\rangle\langle j'||\Psi_{\mu}\rangle\langle\Psi_{\mu}|j\rangle\langle j|e^{in\nu_{\mu}}
\end{equation}
where $U^{n}\Psi_{\mu}=e^{in\nu_{\mu}} \Psi_{\mu}$ has been used.
This is useful in further examining
Equations~\ref{1qubitpert:first} and~\ref{persecond}.

\subsection{Example of a Perturbed Two Qubit Cyclic Network}

For this example we restrict ourselves to an SO(3) cyclic network
because it shows the essential properties for a network undergoing
an acyclic qubit perturbation. Using Equation
\ref{1qubitpert:first} an SO(3) cyclic network perturbed by an
acyclic qubit, such as in Figure~\ref{a}(a), may be examined. In
this case, $U=G$ and the operation of the cyclic network on the
cyclic qubits upon one iteration will be described by
\begin{equation}
  G=G_{up}(\phi)G_{dn}(\phi)
\end{equation}
where $G_{up}(\phi)$ and $G_{dn}(\phi)$ are defined in Equations
~\ref{Gdn} and \ref{Gup} with $\alpha,\beta,\delta = 0$.

For $G^{n}$ iterations the matrix elements are given as a function
of n iterations with the aid of Equation~\ref{matrix:n}.  One sees
from Equation~\ref{G} that it is sufficient to consider the lower
3x3 matrix M.  The matrix elements $m_{j,j'}=m_{j,j'}(n)$ have the
form
\begin{equation}\label{form}
  A(\phi)+B(\phi)cosn\nu_{1}+C(\phi)sinn\nu_{1}
\end{equation}
where the coefficients $A(\phi),B(\phi),C(\phi)$ for each of the
$m_{j,j'}$ are given in Table~\ref{mjj}. (Table~\ref{mjj} uses the
following abbreviations $c=cos\phi$ and $s=sin\phi$.)  Note, that
only the eigenphase $\nu_{1}$ appears in Equation~\ref{form}, this
is due to the fact that $\nu_{0}=\nu_{3}=1$ and $\nu_{2}=-\nu_{1}$
as for all SO(3) matrices.

With the aid of these matrix elements it is now possible to solve
for the evolution of the network after its acyclic qubit
interaction. In order to simplify this example the cyclic network
will initially be placed in one of its eigenstates instead of an
arbitrary superposition as in Equation~\ref{super}. Using
Equation~\ref{eigenstatesG} in section III, the eigenstates of G
are
\begin{gather}
    \Psi_{k} =N_{k}
    \begin{pmatrix}
        0\\
        -sin(\phi)(1-\lambda_{k} cos(\phi))\\
        -sin(\phi)(cos(\phi)-\lambda_{k})\\
        (cos(\phi)-\lambda_{k})^{2}
    \end{pmatrix}
    \begin{matrix}
        |00\rangle\\|01\rangle\\|10\rangle\\|11\rangle
        \end{matrix}
    \quad k=0,1,2 \notag \\
     \Psi_{3} =
    \begin{pmatrix}
        1\\
        0\\
        0\\
        0
        \end{pmatrix}
    \begin{matrix}
        |00\rangle\\|01\rangle\\|10\rangle\\|11\rangle
        \end{matrix}
\end{gather}
with the following normalization factors
\begin{align}
  N_{0}&=\frac{1}{(1-cos\phi)\sqrt{(1-cos\phi)(cos\phi+3)}} \notag \\
  N_{1,2}&=\frac{1}{\sin^2\phi\sqrt{(cos\phi+1)(cos\phi+3)}}\quad
\end{align}
Since this cyclic network has been restricted to the SO(3) group
its eigenvalues can be found from Equation \ref{evalues} with the
trace being set to
\begin{equation}
  TrM=cos^2(\phi)+2cos(\phi)
\end{equation}

In using Equation~\ref{1qubitpert:first} to give the evolution of
the cyclic network after the perturbation, we set the initial
iterations of the cyclic network to $n=0$. This is due to the fact
that the initial state of the cyclic network is in an eigenstate
of $G$ and gives an inconsequential global phase for the $G^n$
iterations prior to the perturbation. Therefore, the evolution of
the cyclic network after the perturbation is given by
\begin{equation}\label{2qubitperturbed1}
  \alpha |0\rangle \otimes e^{in'\nu_{k}} |\Psi_{k}\rangle+ \beta |1\rangle \otimes
  G^{n'}
  \sigma_{x,b} |\Psi_{k}\rangle
\end{equation}
where the final state is found to be entangled into a
superposition of an unperturbed and perturbed cyclic network
state.

The unperturbed part of the final state is $\alpha |0\rangle
\otimes e^{in'\nu_{k}} |\Psi_{k}\rangle $ which evolves according
to the phase $e^{in'\nu_{k}}$.  In contrast, the perturbed part of
the entangled state is $\beta|1\rangle \otimes G^{n'} \sigma_{x,b}
|\Psi_{k}\rangle$ and can be expanded out in the binary basis as
\begin{gather}
  \beta|1\rangle \otimes N_{k}\begin{pmatrix}
        -s(1-c\lambda_{k})\\
        m_{1,2}(n') (c-\lambda_{k})^2-sm_{1,3}(n')(c-\lambda_{k})\\
        m_{2,2}(n') (c-\lambda_{k})^2-sm_{2,3}(n')(c-\lambda_{k})\\
        m_{3,2}(n')
        (c-\lambda_{k})^2-sm_{3,3}(n')(c-\lambda_{k})\\
    \end{pmatrix}
    \begin{matrix}
        |00\rangle\\|01\rangle\\|10\rangle\\|11\rangle
        \end{matrix} \notag \\
    k=0,1,2 \quad
\label{pert-example}
\end{gather}
and
\begin{equation}
   \beta|1\rangle \otimes N_{k}
    \begin{pmatrix}
                 0  \\
        m_{1,1}(n') \\
        m_{2,1}(n') \\
        m_{3,1}(n') \\
    \end{pmatrix}
    \begin{matrix}
        |00\rangle\\|01\rangle\\|10\rangle\\|11\rangle
        \end{matrix}
    \quad k=3
\end{equation}
>From this it can be seen that the coefficient for $|100\rangle$
(leftmost digit belongs to the acyclic qubit) does not evolve in
time with iteration number $n'$. However, the coefficients for the
other three basis states $|101\rangle,|110\rangle$ and
$|111\rangle$ do evolve as a function of iteration number $n'$.
Furthermore, each of the coefficients for the basis states have
the form
\begin{equation}\label{cosfunction:nu}
  A(\phi)+B(\phi)cosn'\nu_{1}+C(\phi)sinn'\nu_{1}
\end{equation}
due to the fact that all of the $m_{j,j'}$ in Table 1 carry this
form. (Note: $A(\phi),B(\phi)$ and $C(\phi)$ for the basis states
may be complex for initial cyclic network eigenstates
corresponding to k=1,2)

\begin{figure}[t]
    \centering{\epsfig{file=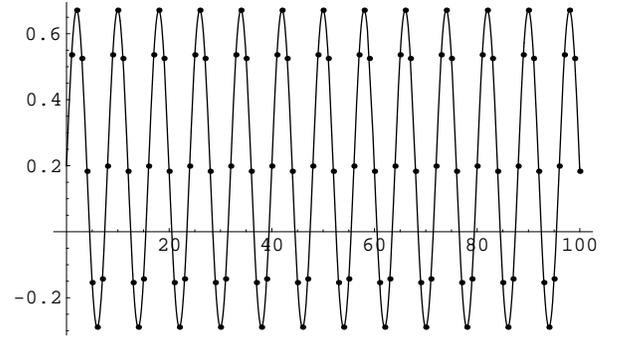,height=1.9in}}
    \caption{The coefficient for the $|110\rangle$ basis state as a function of $n'$ iterations where
    $\nu_{1}=\frac{\pi}{4}$.}
    \label{pi-4}
\end{figure}

As an example of the perturbed evolution consider the case where
the initial eigenstate is set to $k=0$ and the cyclic gate
parameter $\phi$ is chosen so that $\nu_{1}=\frac{\pi}{4}$.  For
this case, the coefficient of the basis state $|110\rangle$ in
Equation~\ref{pert-example} will evolve as shown in
Figure~\ref{pi-4}.  The coefficients for the other two basis
states $|101\rangle$ and $|111\rangle$ are similar in nature.
\begin{figure}[t]
    \centering{\epsfig{file=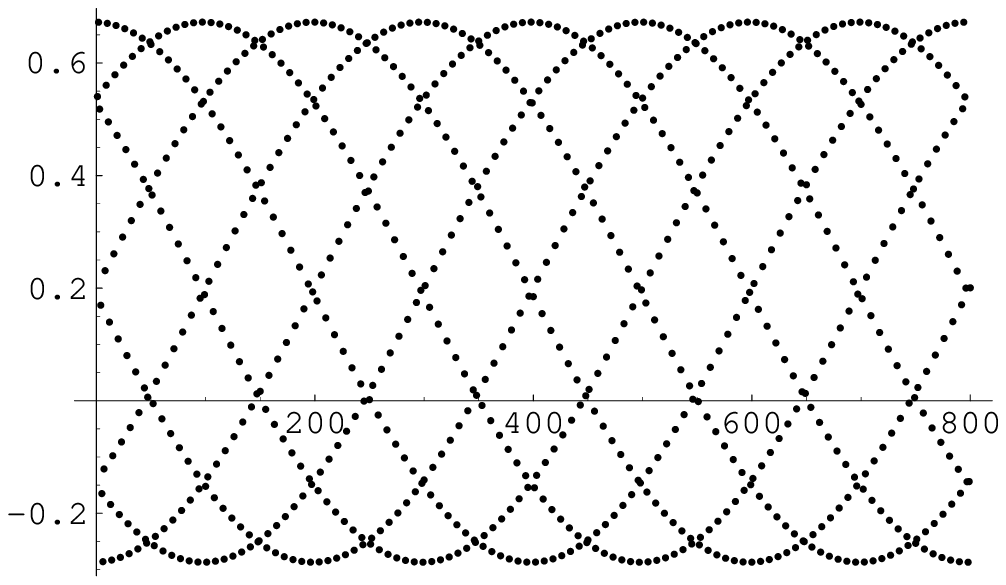,height=1.9in}}
    \caption{The coefficient for the $|110\rangle$ basis state as a function of $n'$ iterations where
    $\nu_{1}=\frac{(\pi+.01\pi)}{4}$.}
    \label{pi-4deviation}
\end{figure}
\begin{figure}[!]
    \centering{\epsfig{file=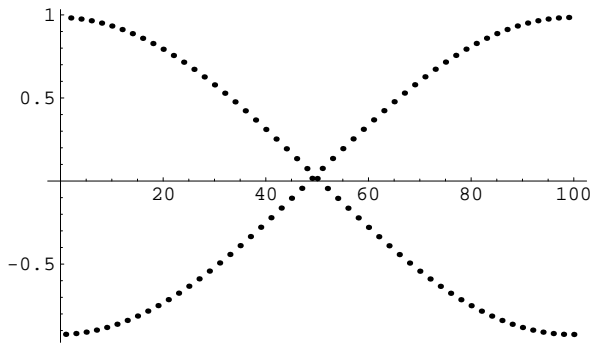,height=1.9in}}
    \caption{The coefficient for the $|110\rangle$ basis state as a function of $n'$ iterations where
    $\nu_{1}=.99\pi$.}
    \label{.99pi}
\end{figure}
The ordinate represents the probability amplitude and the abscissa
corresponds to the time interval $n'$.  The evolution of the
coefficient can be seen on the plot as points on the continuous
curve.  The continuous curve is just a background which represents
Equation~\ref{cosfunction:nu} with $n'$ continuous, and it does
not represent the evolution of the cyclic network's qubits because
the state of the qubits evolve in discrete time steps. It is only
shown here to illustrate how the perturbed cyclic network's
evolution samples the continuous curve and to help explain the
next figure where a small deviation is made to
$\nu_{1}=\frac{\pi}{4}$.

One interesting feature for the evolution corresponding to
$\nu_{1}=\frac{\pi}{4}$ is when a small deviation is made, such as
$\nu_{1}=\frac{(\pi+.01\pi)}{4}$.  For this case,
Figure~\ref{pi-4deviation} illustrates how each of the points in
one period of the continuous curve is modulated by a period of 800
iterations.  In other words, if we choose one of the points in
Figure~\ref{pi-4} and modulate it by the $\frac{.01\pi}{4}$
deviation the point will follow one of the iterated curves in
Figure~\ref{pi-4deviation} for 800 iterations before returning to
its original value.  This may be understood by rewriting
Equation~\ref{cosfunction:nu}, for $n'$ integer, as
\begin{equation}\label{cosfunction:nu:integer}
  A(\phi)+B(\phi)
  cosn'(\frac{\pi}{4}+\frac{.01\pi}{4})+C(\phi)sinn'(\frac{\pi}{4}+\frac{.01\pi}{4}))
\end{equation}
where the factor $\frac{.01\pi}{4}$ modulates the equation and
does not complete a period until $n'=800$.

Another interesting example of this type of modulation is when
$\nu_{1}=.99\pi$ as can be seen in Figure~\ref{.99pi} for basis
state $|110\rangle$. In this case the points are modulated in a
similar manner but the iteration period is $n'=200$. Note, the
period may easily be extended by a factor of 10 by simply choosing
$\nu_{1}=.999\pi$. The fact that the periodicity of these basis
states may be increased in this way may have potential
applications in making cyclic network sensors.

As a final comment for this particular example of a perturbed
cyclic network, it's worth mentioning that for initial cyclic
network states $\Psi_{0},\Psi_{1}$ and $\Psi_{2}$ the perturbed
part of the entangled wavefunction goes into a superposition of
its four eigenstates. However, for an initial state starting in
$\Psi_{3}$ the perturbed part goes into a superposition of three
eigenstates excluding the original $\Psi_{3}$ .  This property
will also be discussed in the next section as a possible
application for a cyclic network memory, or sensor.

\section{Discussion}

The classification and evolution of one and two qubit cyclic
quantum networks for unperturbed and perturbed systems have been
addressed.  Up to now, mainstream quantum information research has
focused on the goal of building a large quantum computer for
factoring or searching.  One possible spin off from this type of
research might be quantum sensors and quantum memories that are
only a few qubits in length.   For this type of quantum
information application, one or two qubit cyclic quantum networks
may be well suited.  This is not only due to the small number of
qubits but also to the simplicity of the repeating gate
operations. The fact that cyclic repetitions may be simpler to
implement experimentally than complicated, varying gate
arrangements, may be an advantage to current experimental efforts.
Other interesting possibilities lie in the exploration of quantum
algorithms and quantum information processing with cyclic quantum
networks.

In the next sections a discussion about these possible research
directions and applications of cyclic quantum networks will
follow.

\subsection{Quantum Algorithms and Quantum Information Processing with Cyclic Quantum Networks}

\begin{figure*}
    \epsfig{file=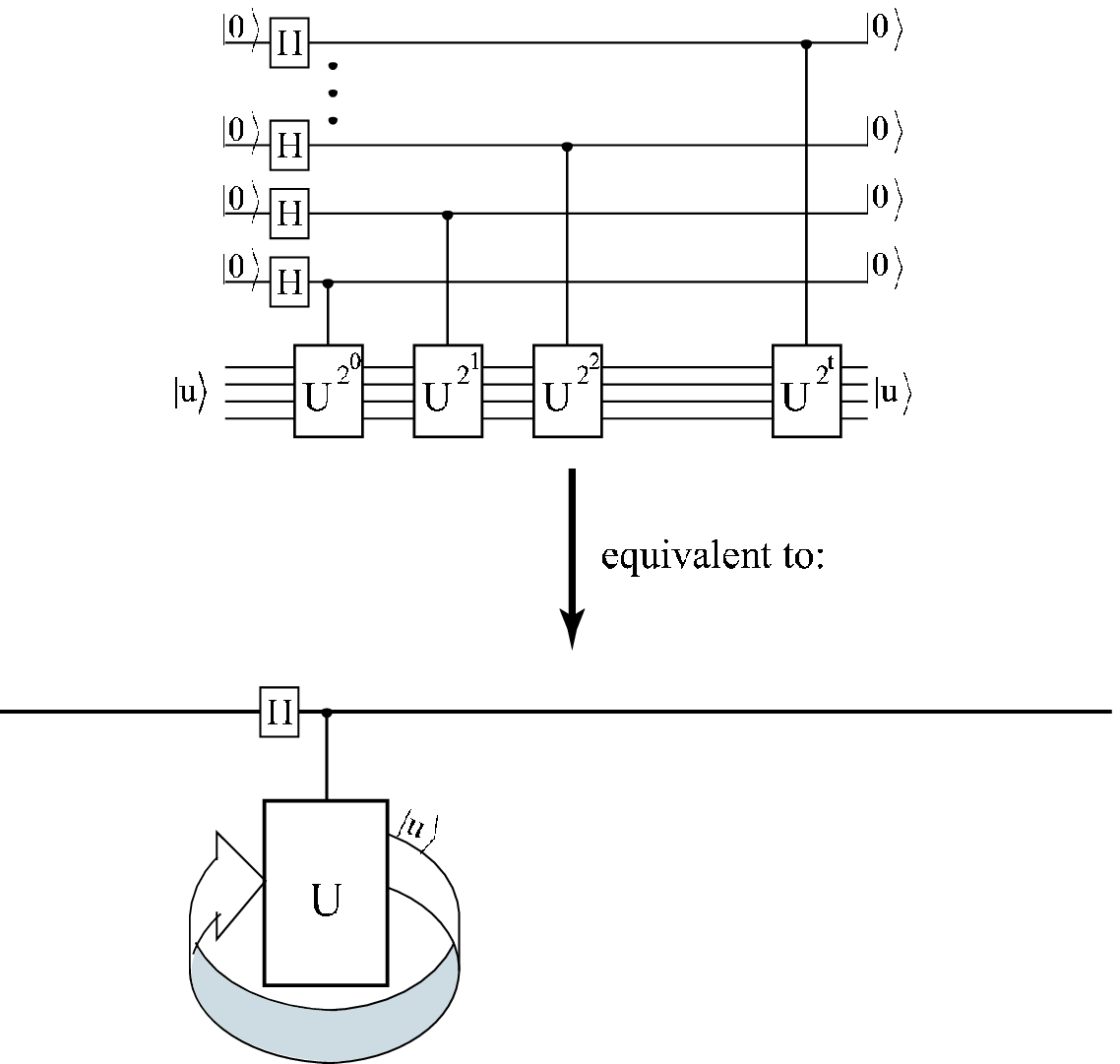}
    \put(-60,301){\makebox(0,0)[bl]{$\scriptstyle{+\thickspace e^{2\pi
        i(2^{t-1}\varphi)}\thickspace|1\rangle}$}}
    \put(-60,271){\makebox(0,0)[bl]{$\scriptstyle{+\thickspace e^{2\pi
        i(2^{2}\varphi)}\thickspace|1\rangle}$}}
    \put(-60,252){\makebox(0,0)[bl]{$\scriptstyle{+\thickspace e^{2\pi
        i(2^{1}\varphi)}\thickspace|1\rangle}$}}
    \put(-60,231){\makebox(0,0)[bl]{$\scriptstyle{+\thickspace e^{2\pi
        i(2^{0}\varphi)}\thickspace|1\rangle}$}}
    \put(-325,107){\makebox(0,0)[bl]{$\scriptstyle{|0\rangle\thickspace
        \ldots \thickspace \otimes |0\rangle \thickspace \otimes |0\rangle \thickspace \otimes |0\rangle}$}}
    \put(-230,107){\makebox(0,0)[bl]{$\scriptstyle{(|0\rangle+\thickspace e^{2\pi
    i(2^{t-1}\varphi)}|1\rangle)\otimes
        \ldots \thickspace (|0\rangle + e^{2\pi
        i(2^{2}\varphi)}|1\rangle)\otimes
         (|0\rangle + e^{2\pi i(2^{1}\varphi)}|1\rangle)\otimes (|0\rangle + e^{2\pi i(2^{0}\varphi)}|1\rangle)}$}}

    \caption{The acyclic network at the top is the phase estimation network which is equivalent to the cyclic network
    shown below it.  The operator U has eigenstate $|u\rangle$ with corresponding eigenvalue $e^{2\pi i \phi}$. Note, the
    cyclic network must evolve for $U^{2^{t}}$ times for the $t$-th acyclic qubit crossing
    the acyclic control line.  The single qubit operator H puts $|0\rangle$ into an equal superposition
    $\frac{1}{\sqrt{2}}(||0\rangle+|1\rangle)$ where the $\frac{1}{\sqrt{2}}$ terms have been left off the diagram.}
    \label{cyclic:phase-estimation}
\end{figure*}

As mentioned earlier, the quantum phase estimation algorithm can
easily be expressed more compactly with cyclic quantum networks.
For instance, Figure~\ref{cyclic:phase-estimation} shows the phase
estimation algorithm as typically depicted with an acyclic
array~\cite{revisited}.
It also shows the equivalent cyclic network with the acyclic
qubits acting in a conditional manner on the cyclic network's
unitary operation.

Although expressing the quantum phase estimation algorithm in this
way is quite trivial and suggests no new algorithms, it does serve
as an example of a compact, cyclic network with an acyclic qubit
line (similar to the ones investigated in the last section).
Another example of an algorithm that can be compactly expressed in
terms of cyclic networks is Grover's Search
Algorithm~\cite{grover}. This is due to the repeating unitary
operation sometimes called Grover's iterate which can also be
modelled by a cyclic network. This common pattern of iterative
operations in quantum algorithms is reminiscent of the observation
that quantum algorithms resemble a multiparticle
interferometer~\cite{revisited}.  This observation (quantum
algorithms resembling multiparticle interferometers) supports the
finding that many quantum algorithms including Shor's algorithm
may be viewed as a phase estimation process.  It is possible that
continued investigations into the iterative nature of these
algorithms viewed in terms of cyclic networks may bring new ideas
into quantum algorithm design.

Aside from expressing quantum algorithms with cyclic networks, one
possible scope of applications for simple cyclic networks might
involve methods of connecting them to existing acyclic arrays.
These type of connections could potentially act as a type of
subroutine for the acyclic arrays, in that the known perturbed
evolution for the cyclic networks might serve as a type of module
algorithm within the whole array.

Other applications may emerge from finding ways of connecting
these simple cyclic networks together in chains so as to
understand the evolution of the overall composite network from the
known state-evolution of the simpler networks.  In other words,
the evolution of qubits belonging to a group U$(2^q)$ may be
understood by simply writing down the wavefunction for the system
which is comprised of the wavefunctions for the simpler cyclic
networks.  This type of analysis could potentially bring on new
ways of understanding quantum information processing for large
$2^q$ dimensional Hilbert Spaces.

\begin{figure}
    \centering{\epsfig{file=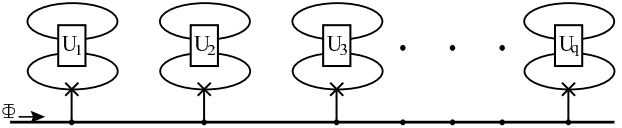,height=0.65in}}
    \caption{A chain network comprised of two qubit cyclic networks linked by one acyclic line.}
    \label{chain:fig}
\end{figure}

One example of a composite cyclic network belonging to the
U$(2^q)$ group can be seen in the chain network of
Figure~\ref{chain:fig}.
This network obviously gives a highly entangled number of qubits.
The total network is arranged so that each cyclic network's qubits
move around the loops in one time step and cross the control gates
in unison.  Also, only one time step (iteration of qubit) occurs
for the total network when the acyclic qubit moves from
interacting with one cyclic network to interacting with the
neighboring network. The total network's qubit evolution after the
acyclic qubit $\Phi$ passes all the cyclic networks in q steps is
given by
\begin{multline}\label{chain-equation}
\{ P_{0} \otimes [ U_{q}^{(n'+q)}\otimes \ldots \otimes
U_{3}^{(n'+q)} \otimes
 U_{2}^{(n'+q)}\otimes U_{1}^{(n'+q)}] \\
 + P_{1} \otimes U_{q}^{n'} \otimes \ldots \otimes U_{3}^{n'}
\otimes U_{2}^{n'} \otimes U_{1}^{n'} [\sigma_{x,b}^{q}U_{q}^{q}
\otimes \ldots \otimes U_{3}^{(q-2)}
\sigma_{x,b}^{3} U_{3}^{2} \\
\otimes U_{2}^{(q-1)}\sigma_{x,b}^{2} U_{2}^{1} \otimes U_{1}^{q}
\sigma_{x,b}^1] \}\\
|\Phi \otimes \psi_{q}
\otimes\ldots\otimes\psi_{3}\otimes\psi_{2}\otimes\psi_{1}\rangle
\end{multline}
where the superscripts in $\sigma_{x,b}$ denote the cyclic network
being operated on, and $n'$ corresponds to the number of
iterations after the perturbation of the cyclic networks. Note,
the $P_{0}$ and $P_{1}$ operate on the acyclic qubit.

As a final remark on compact cyclic networks, it should be
mentioned that chains of cyclic networks may be useful in the
field of neural networks. For instance, by connecting just a few
input and output lines to a cyclic network (or a chained set of
cyclic networks), one can begin to ask questions on how this
network might be trained in the same sense that the highly
cyclical neural networks are trained~\cite{neural1, neural2}.  One
simple method might rely on making measurements on the output
qubits followed by a feedback of adjustments to the control gate
parameters.  The network can then be re-tested to see how the
changes affect a new set of inputs.  Simple procedures of this
type using cyclic networks of quantum gates may be worth exploring
in that it may benefit research in the field of neural networks.

\subsection{Quantum Memories}

One of the original motivations for studying cyclic quantum
networks was the fact that quantum algorithms typically mention
quantum memories, but give only one way of achieving this type of
memory.  This method implicitly relies on a system where no
quantum operations act on a set of qubits (or a quantum register)
until it is needed.  Therefore, the quantum information in a set
of qubits can be retained by sheltering the qubits from any
interactions that may cause the information to be changed or lost.

An alternative to this type of quantum memory will be suggested in
this section, but before doing this, it is helpful to re-cast the
typical quantum memory into cyclic quantum circuit language. This
is not difficult to do because it only involves setting the qubits
(or register) into a cyclic network where the only operation being
applied is the identity operation.  In other words, the cyclic
operators in Figures~\ref{twoqubit-gates} are replaced by an
identity operator whereby the qubits cycle about in an unaffected
fashion.

In section III a discussion of all the one and two qubit groups of
operators and their corresponding eigenstates and eigenvalues was
given. These cyclic networks can potentially be used as quantum
memories due to the fact that a set of qubits in an eigenstate of
the cyclic network is unaltered (except for an inconsequential
global phase) upon passing through the gates.  Therefore, quantum
information in the form of an eigenstate may be stored in a cyclic
network.

However, if an arbitrary state expanded over the eigenstates of
the cyclic network undergoes iterations, then the phases do become
relevant. In this case, the differences of phases between the
eigenstates act as interference terms in the binary basis altering
the initial state of the memory. Nevertheless, the initial state
may be recovered (in a reversible sense) due to the fact that the
number of iterations $n$ that the cyclic network has evolved since
the memory was saved is known and that the unitary operation U for
which the cyclic network operates on the qubits for one iteration
is also known.

\begin{figure}
    \epsfig{file=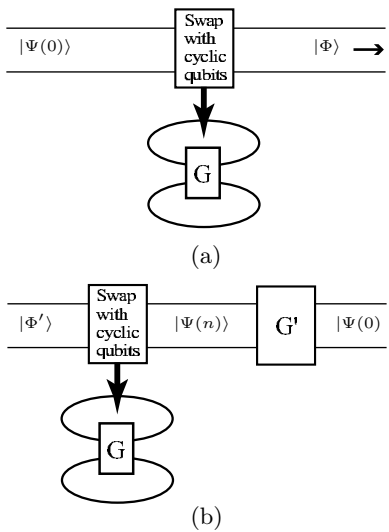}
       \put(-140,170){\makebox(0,0)[bl]{$\scriptstyle{ |\Psi(0)\rangle
       \hspace{32.0mm}|\Phi\rangle}$}}
       \put(-75,89){\makebox(0,0)[bl]{\small{(a)}}}
       \put(-140,65){\makebox(0,0)[bl]{$\scriptstyle{|\Phi'\rangle \hspace{16.0mm} |\Psi(n)\rangle \hspace{14.0mm} |\Psi(0)}$}}
       \put(-75,-10){\makebox(0,0)[bl]{\small{(b)}}}
    \caption{(a)- Two qubits in an unknown quantum state $|\Psi(0)\rangle$ are placed into the cyclic quantum memory by
     swapping with two qubits in an arbitrary state $|\Phi\rangle$ initially in the cyclic network G. (b)- After saving the
     unknown quantum state for n iterations, two new qubits
     in arbitrary state $|\Phi'\rangle$ are swapped with the evolved, quantum state $|\Psi(n)\rangle$.  The state
     is returned to its initial value by applying the inverse operation
     $G'$.}
    \label{Quantum-Memory}
\end{figure}

Consider the scheme described in Figure~\ref{Quantum-Memory}(a).
In this description of a cyclic quantum memory, two qubits in an
unknown quantum state $|\Psi(0)\rangle$ are saved in the cyclic
quantum memory by swapping with the two qubits in the cyclic
network G for which the parameters $m_{j,j'}$ in G are known. The
unknown quantum state is allowed to evolve n iterations
corresponding to the time the unknown quantum state is saved in
the cyclic memory.  After n iterations the unknown quantum state
may be retrieved as in Figure~\ref{Quantum-Memory}(b) by bringing
two new qubits in the state $|\Phi'\rangle$ and swapping with the
evolved, unknown quantum state $|\Psi(n)\rangle$ in the cyclic
network . The state $|\Psi(n)\rangle$ may be reversibly set back
to its initial state $|\Psi(0)\rangle$ by applying the operation
$(G^{\dag})^{n}$. Interestingly, this does not require n
iterations of $G^{\dag}$ due to the fact that
Equation~\ref{matrix:n} may be used to generate the matrix
elements to $G'=(G^{\dag})^{n}$. Therefore, by applying the single
operation $G'$, the initial unknown quantum state may be
retrieved.

In summary, this scheme allows cyclic networks to implement a
novel quantum memory, but whether or not this type of quantum
memory is of any use in quantum algorithm design or implementation
remains an open question.

\subsection{Quantum Sensors}

In the context of cyclic quantum gate networks perturbed by an
acyclic qubit, a quantum sensor may be defined.  This is not
difficult to achieve since a quantum sensor is nothing more than a
recorder of an interaction.  Furthermore, since the last section
discussed methods of implementing a quantum memory via cyclic
quantum gate networks, the quantum sensor can be thought of as a
type of quantum memory recording a qubit interaction with the
cyclic network.

One specific example of a quantum sensor that may be analyzed is
the example of a perturbed two qubit cyclic network in section
IV-B. If the goal of the quantum sensor is to be able to detect
the passage of a qubit in state $|1\rangle$ on the acyclic line,
then a measurement of the cyclic network's qubit states can give
this information if the cyclic network's evolution is perturbed
from its unperturbed evolution.

One simple case for this is the cyclic network with initial state
chosen to be $\Psi_{3}$. If an acyclic qubit state $|1\rangle$
interacts with the cyclic network via a Control-Not gate, the
cyclic network goes into a perturbed superposition state of
$\Psi_{0},\Psi_{1},\Psi_{2}$ and excludes the initial state
$\Psi_{3}$ as mentioned in Section IV-B. A simple measurement to
detect $\Psi_{3}$ can determine whether or not an acyclic qubit
state $|1\rangle$ has interacted with the cyclic network.

This example on its own is not very impressive since a simple
Control-Not gate can give this same type of information without
having to make use of a cyclic network.  However, a quantum sensor
as a perturbed cyclic quantum network gives a method of recording
an interaction where the recording qubits undergo repeated unitary
operations. This type of sensor may be useful in experimental
situations where it is simpler to have repeated gate operations.

Another reason that cyclic quantum sensors may be of use, is in
the possibility that quantum algorithms may be applied to the
sensing process and thereby increasing capabilities beyond that of
"classical" devices. This however remains to be investigated.

\section{Conclusion}

The structure of one and two qubit cyclic quantum gate networks
have been classified.  The unperturbed evolution for these
networks has been addressed, and a specific class of perturbations
have been examined. A discussion on the potential aspects of these
networks in regards to new directions in algorithm design with
cyclic quantum networks has also been given.

One specific new finding is a novel implementation of a quantum
memory using cyclic quantum networks. A quantum memory via cyclic
networks can potentially be used in experimental systems where
repeating unitary operations are preferred to traditional quantum
memories where information is preserved via no applications of
quantum gate operations. Also, a type of quantum sensor similar to
the cyclic quantum memory (modified with an acyclic perturbation
line) has been given.  One possible research direction for cyclic
quantum sensors will be to find ways of increasing sensor
capabilities with quantum algorithms.

\section{Acknowledgment}
We would like to thank Alberto Rojo, Ben Zeidman and Roy Clarke
for helping the first author complete this work as partial
fulfillment of the requirements for the Doctor of Philosophy in
Applied Physics at the University of Michigan. This work was
facilitated in part by a National Physical Science Consortium
Fellowship and by stipend support from the National Security
Agency.  Additional support for this work was provided by the U.S.
Department of Energy, Nuclear Physics Division, under Contract No.
W-31-109-ENG-38.

\bibliography{PcJournal_bib}% Produces the bibliography via BibTeX.

\end{document}